# From Trust to Truth:
# Actionable policies for the use of AI in fact-checking in Germany and Ukraine

Practical recommendations for AI and fact-checking communities and funding bodies in Ukraine and Germany regarding the most promising areas for developing AI-based tools for fact-checking and disinformation monitoring, testing frameworks, necessary funding and media regulation.

**Veronika Solopova**
**Technische Universität Berlin**

**29 November 2024**

# Plan

# Executive Summary

The rise of Artificial Intelligence (AI) presents unprecedented opportunities and challenges for journalism, fact-checking and media regulation. While AI offers tools to combat disinformation and enhance media practices, its unregulated use and associated risks necessitate clear policies and collaborative efforts.

## Purpose

This policy paper explores the implications of artificial intelligence (AI) for journalism and fact-checking, with a focus on addressing disinformation and fostering responsible AI integration. Using Germany and Ukraine as key case studies, it identifies the challenges posed by disinformation, proposes regulatory and funding strategies, and outlines technical standards to enhance AI adoption in media. The paper offers actionable recommendations to ensure AI's responsible and effective integration into media ecosystems.

## Opportunities

AI presents significant opportunities to combat disinformation and enhance journalistic practices. However, its implementation lacks cohesive regulation, leading to risks such as bias, transparency issues, and over-reliance on automated systems. In Ukraine, establishing an independent media regulatory framework adapted to its governance is crucial, while Germany can act as a leader in advancing EU-wide collaborations and standards. Together, these efforts can shape a robust AI-driven media ecosystem that promotes accuracy and trust.

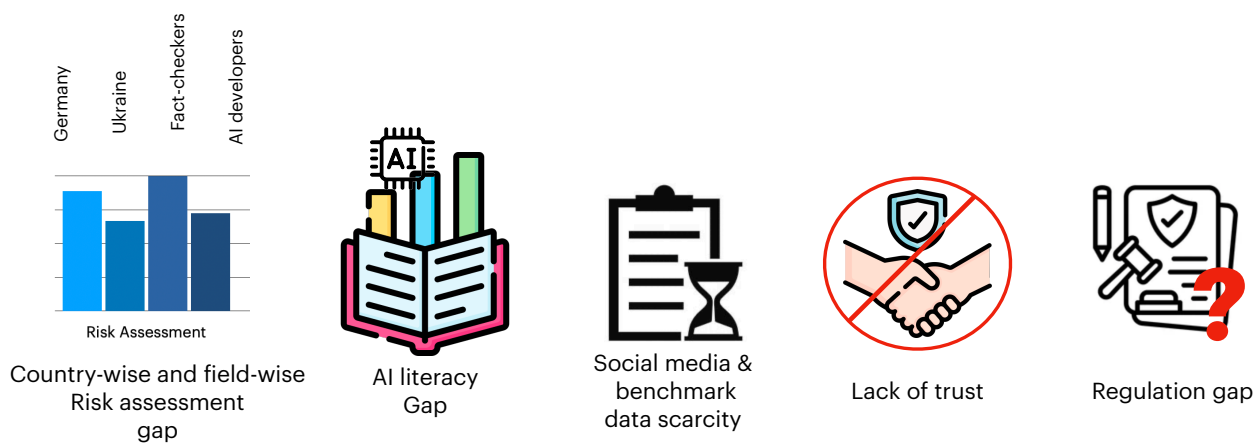

Country-wise and field-wise Risk assessment gap | AI literacy Gap | Social media & benchmark data scarcity | Lack of trust | Regulation gap

## Key Findings

Interviews with experts from Germany and Ukraine revealed the following key insights about the state of AI in journalism and fact-checking:

1. **Country and Specialisation gaps.** Quantitative survey results point to the difference between risk perception of, on the one hand, German and Ukrainian communities and, on the other, AI and fact-checking ones.

2. **AI Literacy Gaps.** Many policymakers and journalists lack sufficient training in AI tools, creating barriers to effective implementation and oversight.
3. **Data Scarcity.** Access to high-quality, cross-platform data remains the primary bottleneck for AI research and development, exacerbated by restrictive platform policies.
4. **Trust Issues.** Fact-checkers and journalists are sceptical about AI's transparency and reliability, often perceiving it as a "black box" solution, also pressured by the regulations against experimenting with AI tools.
5. **Regulation gaps.** While Ukrainian law and governance generally lack comprehensive media law and should focus on the creation of regional media regulators, both country do not have any comprehensive regulation for the usage of AI in media and fact-checking, which leaves a place for cautious interpretation and fears.

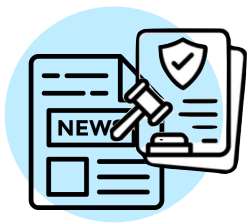
Media regulation

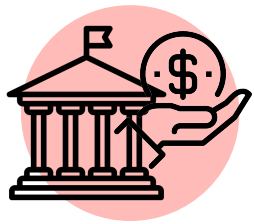
Governmental funding

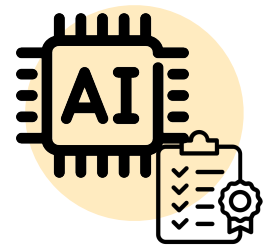
AI developments standards

---

## Recommendations

I. **Media Regulation**:

   A. Ukraine should establish independent regional media regulatory bodies to oversee AI integration into journalism, following the principles of transparency, independence from government control, and alignment with EU frameworks such as Germany's Medienstaatsvertrag and the Digital Services Act (DSA).
   B. Germany and the EU as a whole should strengthen disinformation regulation by fostering public consensus on freedoms of speech in the context of disinformation campaigns, updating national legislation to effectively and directly counter modern informational threats, also aligning with EU standards on this matter.
   C. Clear Media Regulation for AI in fact-checking in both Germany and Ukraine should be established. This regulation should emphasise AI as a supportive tool rather than a replacement for human journalists, ensuring accountability through multi-level oversight and transparent decision-making processes.

II. **Governmental Funding Strategies**:

   A. Increased funding for AI literacy programs is essential to empower journalism, government, and education stakeholders with a clear understanding of AI's capabilities and limitations.
   B. Investment in interdisciplinary and technical academic programs targeted at students with backgrounds in Law and Administration as well as fully funded PhD digital journalism and digital humanities positions, will foster the next generation of AI-policy experts.
   C. Enhanced collaboration between EU member states and candidate countries like Ukraine and Moldova can strengthen regional resilience to disinformation and support innovation in AI-driven fact-checking.

D. Create a consortium of researchers, regulators and social media platform representatives working towards the creation of the European Association for Social Media Research which would address data and computational resources scarcity. This consortium would build infrastructure for simplified research access to real-time social media data sharing, community datasets, and further supervised access to platform data under the Digital Services Act (DSA) with vetted membership, anonymisation and ethical oversight ensuring data privacy. The association would enable cross-border collaboration, including researchers from non-EU countries like Ukraine, and provide shared computational infrastructure to reduce costs and environmental impact. It would attract international projects, support innovation, and train experts in AI, AI-driven content moderation and fact-checking through a specialized post-graduate program.

**III. AI Development Standards**:

A. AI tools must prioritise explainability, reliability, and scalability and go paired with user-friendly explainable AI (XAI) interpretation techniques.
B. An iterative 6-step testing process is proposed, including multiple out-of-distribution datasets, counterfactual and fairness analyses and tests with end-users, which are critical for aligning tools with the needs of journalists and fact-checkers and providing the fact-checkers with the feeling of ownership and collaboratively achieved performance progress.
C. Prioritise workflows, which prevent the biasing of fact-checkers with AI tools and are naturally integrated with existing teamwork workflow.
D. Focused investment in benchmarking datasets, data-sharing frameworks, and computational resources will address existing bottlenecks and enable the development of robust AI solutions, also in the context of the policy recommendation II.D.

A list of possible **success metrics** is proposed, including:
- The volume of data accessed under DSA.
- The number of cross-border collaborations established.
- Research projects and publications generated.
- Professionals trained in AI-driven moderation.
- Reduction in content moderation errors and fines.
- Increased public trust in AI fact-checking tools,
- others.

**Timeline**

Achieving these goals requires a multi-phase approach over the next five years. Immediate steps include establishing foundational regulations, funding AI literacy programs and establishing the consortium working towards the Association of European Social Media Research, which will entail long-term investments in shared infrastructure, collaboration with social media platforms, and data-sharing frameworks will ensure sustainable progress.

# Introduction

According to the American historian of Eastern Europe, Timothy Snyder's "On Freedom" (Snyder, 2024), freedom of Speech is about creating free speakers, not about digital oligarchs as the latter want it to appear. One of the main goals of the disinformation campaigns is to create an atmosphere of confusion and moral relativism. As a society, we should remember that we can and should discuss values and norms, but we should all respect facts and that these exist. Social media is set up for disinformation, and it does not seem to be a bug, it is a feature. It is meant to drive us into emotional states which are anti-democratic and get us to do things which are not in our own best interests. Our main goal in these post-truth times is to support institutions that generate truth and facts: universities, investigative journalists, reporters, and fact-checkers.

The pervasive crisis of trust in global news has prompted widespread debate on measures to restore credibility (Flew et al., 2020; Gaziano, 1988). With the increasing reliance on digital news, especially among younger generations, and the trend of avoiding coverage of issues like COVID-19 and the Russian-Ukrainian conflict due to perceived negativity and low credibility (Coster, 2022), digital platforms have come under greater scrutiny. Political news, in particular, is viewed with suspicion, often seen as agenda-driven and potentially containing propaganda (Mont'Alverne et al., 2022; Flanagin and Metzger, 2000; Kalogeropoulos et al., 2019). A prominent concern is the rise of state-sponsored Kremlin propaganda, as only a small fraction of Russian bots are being detected, while pro-Russian activity on platforms like Twitter (now X) continues to proliferate, amassing substantial engagement and influencing public perception (Geissler et al., 2023). Many European citizens are targeted by Kremlin propaganda campaigns designed to reduce public support for Ukraine, cultivate mistrust and disunity, anti-American sentiment and influence elections (Meister, 2022).

Starting around 2010-2014, to counteract the new threat to democratic societies, big media outlets all around the world expanded their research departments to scale fact-checking capacities, while specialised fact-checking NGO organisations emerged and raised to the new challenges, e.g. CORRECTIV, Les Décodeurs, Demagog were founded in 2014; AFP Fact Check, maldita.es - in 2017-2018. With the ever-growing numbers of disinformation posts and disinformation narratives (Allcott and Gentzkow, 2017), fact-checking agencies and units started multiple collaborative projects with research institutions to automate some part of the fact-checking work.

Developing AI for fact-checking faces multiple challenges, especially given the dynamic and nuanced nature of disinformation. Rapidly evolving tactics in disinformation (Adriani, 2019), along with culturally embedded nuances—make it difficult for AI models to remain effective without constant updates and retraining (Horne et al., 2020; Solopova et al., 2023; Bozarth & Budak, 2020). Context understanding is another obstacle, as disinformation often relies on subtle cultural references or manipulations that current Natural Language Processing (NLP) models struggle to interpret accurately (Zhou et al., 2023). Additionally, distinguishing between reliable and dubious sources without embedding AI biases is critical, as biases can undermine user trust. However, achieving this level of reliability is challenging because data distributions shift quickly, and obtaining timely, high-quality data for training and adaptation is notoriously difficult. Moreover, cultural perceptions can polarise factual interpretation, such as differing views on historical events, which makes it hard for AI to provide universally accepted conclusions.

Meanwhile, AI regulation is also a rising field. While the EU can boast of the most progressive AI law in the West, it still fails to capture all of the complexity of the changing digital landscape and balance the exploratory freedoms with regulations for safety, transparency and data privacy concerns.

Germany and Ukraine, which this study focuses on, provide two distinct and intertwined case studies in the fight against disinformation, each with challenges reflecting broader European and global concerns. Germany, as one of the EU's most influential member states, has taken on a critical role in shaping policy responses to disinformation within the Union. Germany's successful role as a key incubator for the EU's regulatory approach to digital technology is not widely appreciated or understood at home, and digitalisation is perceived as slow and inert (Hagebölling et al. 2022).

Germany's deep-seated commitment to privacy and data protection, while pivotal, often necessitates a cautious approach to adopting new technologies for AI fact-checking, as regulators strive to strike a balance between innovation and security. In contrast, Ukraine faces an entirely different set of obstacles. As a young democracy, Ukraine's media landscape is still developing, with limited well-established and independent media regulation frameworks in place (Korbut, 2021; Dzholos, 2024). The country's unique experience as the frontline of hybrid warfare with Russia has also shaped its digital environment, where disinformation is weaponized to weaken public morale and international support. Despite these challenges, Ukraine has rapidly adopted digital solutions to counter disinformation as part of its broader wartime resilience. Yet, issues of media independence, particularly in the regulation of content and AI transparency, reveal the pressing need for adaptable fact-checking frameworks that can navigate both ongoing conflict and Ukraine's democratic development.

This paper draws on interviews with German and Ukrainian stakeholders from fact-checking, AI development, regulatory bodies, and government. It examines the current level of understanding between these groups in Germany and Ukraine, aiming to identify policy recommendations that could help reduce AI scepticism in fact-checking and government in both countries. The paper also suggests potential funding and regulatory initiatives to strategically address these challenges, while fostering opportunities for mutual exchange and learning—recognizing that Germany and Ukraine each have valuable insights to offer the other.

Our interviews revealed that fact-checkers from both countries in our survey already use a wide range of AI-enabled tools in explicit and implicit ways. Although these are not the tools that detect the presence of disinformation in texts directly, in fact, in addition to conventional spellcheckers, automatic subtitling, and automated interview transcription, most of the fact-checking teams we surveyed focus on the tools for detection of generated and manipulated content.

One of the most used applications is *Inverse Image Search* provided by various search engines (Google, Bing, Yahoo, etc) which helps investigators identify where else in the Internet such image has been used, which is especially useful in identifying false context usages and geolocation. Invid, a browser extension that helps with photo forensics, like verifying the authenticity of online videos and images, is a tool mentioned by German and Ukrainian journalists. The Deutsche Welle team reported the high performance of Truemedia[1] especially for generated and manipulated images. The VoxUkraine team also acknowledged trying out Face search tools, although they expressed concerns about the potential misuse of such tools.

Some media outlets, like the Rundfunk Berlin-Brandenburg (RBB) research service team, focused on testing the tools for future threats, so to speak, identifying tools which may be useful in the future with the development of the technology. For example, they are looking into the tools verifying exact geolocation, which, as of now, can successfully identify a city but struggles with more localised identification and deepfake detection. Fact-checker respondents from both countries confirm that currently, human analysts identify deepfakes better and quicker than AI tools, but all of them expressed concerns about the future, as the deepfake technology gets better and more accessible. Meanwhile, most of the respondents' teams already actively looked into ChatGPT's usefulness in summarising longer reports or identifying the place in the document where they could find the information they searched for.

---

[1] https://www.truemedia.org/

Moreover, over the last years, most of the news organisations we surveyed initiated and participated in various partnerships with AI research groups at universities and companies with fascinating, high-risk ideas. For example, Ukrainian Detectormedia is working with developers on a database to represent the network of media collaborators on Ukrainian-occupied territories. Deutsche Welle (DW) worked on Invid and Digger[2] for deepfake detection in partnership with Fraunhofer Institute, Athens Technology Center and many others. VoxUkraine worked on an initiative together with Mantis Analytics for automated multilingual propaganda detection based on their Propaganda Diary[3] data. CORRECTIV is in partnership with Technische Universität Berlin (TUB) MTEC lab and Ruhr University Bochum Linguistics Forensics Department, working on AI tools for CORRECTIV.Faktenforum, which is an AI-empowered platform for citizen journalists, is united in a desire to learn to fact-check and contribute to the democratic processes with community fact-checks curated by professional journalists. Another impressive TUB and *Deutsches Forschungszentrum für Künstliche Intelligenz* (DFKI)-partnered "news-polygraph" project in collaboration with DW and RBB, develops a technology platform that uses AI to identify targeted disinformation ("fake news") and combines various tools for recognising manipulations of images, sound, videos and texts. These projects are only a few examples of such impressive initiatives funded in recent years, with dozens more in the making or at their closing stages, being funded over multiple different Federal Ministry of Education and Research (*Bundesministerium für Bildung und Forschung,* BMBF)) funding clusters, German Research Foundation (*Deutsche Forschungsgemeinschaft,* DFG), and European Union projects.

However, if an important measurement of success for such an initiative would be the full implementation of the project results into the application partner's fact-checking routines, resulting in increased efficiency and better fact-checking experience, this does not always seem to be the case. In many a project, few of the tools outlined in the project proposal stage, at the end make it into the final product and fact-checking workflow, while fact-checkers also seem unaware of the current status of development. For instance, DW reported that with the Digger project running for at least 3-4 years, they are not yet satisfied with the resulting tool and cannot fully implement it into their fact-checking routines. Moreover, none of the older tools produced as a result of such partnerships was mentioned as a part of the media teams' fact-checking toolkit in our interviews. The exception here is **Invid[4],** developed by a massive cross-country consortium of research, industry and media partners, including CERTH, Modul Technology GmbH, Universitat de Lleida, Exo Makina, WebLyzard Technology GmbH, Condat AG, APA-IT Informations Technologie GmbH, Agence France-Presse and Deutsche Welle. However, one of the few respondents who did not mention this tool was precisely the Deutsche Welle representative.

In the case of Ukrainian projects, after a series of iterations in an attempt to articulate and formalise a task based on VoxCheck's database of propaganda fact-checks, the VoxUkraine fact-checking team realised they were unable to identify a task they would be ready to delegate to an AI agent. Detectormedia's collaborators' network was mentioned as a long-term project, with the usefulness of the outcomes yet to be proved, especially in a dynamically changing landscape of war.
These cases uncover various problems in communication, collaboration structure, project outlining, testing and funding incentives.

---

[2] https://digger-project.com/)

[3] https://russiandisinfo.voxukraine.org/en

[4] https://www.invid-project.eu/

# Background and Context

## Fact-Checkers' Main Concerns

*The next sections are based on interviews with media, government and AI developers' representatives from Ukraine and Germany. A full list of interviewees can be found in the Appendix.*

Our fact-checking respondents from various organizations in Ukraine and Germany expressed significant concerns about the current limitations of AI tools for fact-checking, emphasizing the need for human oversight and careful ethical considerations.

Trust and Reliability concerns are shared among representatives of both countries. Ukrainian fact-checkers highlighted that fact-checking requires a level of critical thinking and contextual awareness that AI lacks. In their experience, models like GPT often provide incorrect sources or draw inaccurate conclusions—a phenomenon known as "hallucination." It also often favours sources that confirm user hypotheses, which can lead to **cherry-picking** and reinforce existing biases and be a serious problem for an objective investigation. VoxUkraine has also observed a notable imbalance in AI's error patterns, with **false negatives** being a more prominent problem. This can be said for both relevant document retrieval and generated content detection: while flagged content is usually flagged correctly, AI tools often overlook falsehoods that the human eye would easily catch. Furthermore, while AI tools perform reasonably well with image verification, they require source images and show a significant drop in performance when screenshots are used, which seriously complicates the investigation. AI models also struggle with video content more than with images. Overall, human analysts are often needed to verify complex or nuanced cases. This means that easy cases, which would anyway not take much human effort to fact-check, are identified well, while difficult cases, where human fact-checkers could use help, are also almost impossible tasks for the AI.

German fact-checkers also see AI as a supportive tool rather than a decision-maker. While AI can assist in gathering and sorting data, final judgments rest with human reviewers. Similarly to Ukrainian colleagues, RBB also has had mixed experiences with GPT-based tools in text analysis. While it can sometimes provide helpful information, it frequently fabricates source details like page numbers and tends to deliver incomplete or imprecise information. According to them, AI could be useful for large-scale investigative projects, where it can narrow down datasets, but these projects still require careful human review to ensure accuracy and validity.

At Deutsche Welle, the reliability of AI detection tools is also a major concern. Despite recent advances, tools for detecting manipulated media or deepfakes remain inconsistent, especially in **audio verification**. DW has rigorously tested AI in this area, but findings show that current tools lack the reliability needed for accurate fact-checking. In image detection, where AI is often tasked with identifying signs of manipulation, the results tend to be inconclusive, offering only probabilities rather than clear indicators of manipulation. For fact-checkers who require concrete evidence, this level of ambiguity is insufficient. Geolocation is another area where AI struggles, identifying the countries but being unable to correctly narrow down the city.

Comparing a more distinct focus of the fact-checking communities of these two countries, in contrast to German colleagues, Ukrainian fact-checkers focused on cyber security and adversarial attack concerns. They highlighted that AI tools can be vulnerable to hacking or other forms of cyber manipulation, posing serious security concerns.

German media representatives strongly emphasise ethical and legal standards—especially in view of the *Sorgfaltspflicht* (en. diligence obligation) that governs media work in Germany and mandates that media organisations carefully trace and justify their conclusions. They prioritise **traceability, transparency, corporate data privacy, and compliance with established review processes.**

The reputation of the fact-checkers is hard won and easy to lose, even with one misleading fact-check out of thousands of correct ones. Still, they happen. That is why, according to CORRECTIV, for fact-checkers, it is important for the readers to trust their processes and standards even more than their conclusions, knowing that even if a mistake may happen, it is made clear and transparent what went wrong and that it will be corrected.

For instance, CORRECTIV, to meet these standards, emphasises that AI-reviewed content should also be reviewed by human evaluators, traditionally following a "six-eye principle," (de. *Sechs-Augen-Prinzip*) where multiple reviewers independently verify AI-driven findings. This raises important questions about the number of reviewers needed to uphold accuracy in AI-supported workflows: can AI be considered the 3rd pair of eyes or only 4th, in which case its claimed efficiency boosting capacity for humans would be yet to be proved. Another concern they expressed, in terms of LLM usage, is the lack of transparency in terms of how OpenAI will use the input data in the future. Therefore, their fact-checkers are not allowed to put investigation materials into it.

Finally, all of the German media representatives agree that AI-driven conclusions must be understandable and verifiable to allow fact-checkers to respond to public inquiries about the source and accuracy of their findings. For instance, highlighting the exact part of the video or images with traces of manipulation of generation artefacts helps journalists to quickly check the concerned frame or part of the image and use it as evidence for the fact-check.

## Quantitative study

During the interviews, we asked the respondents to tell which minimal accuracy from 0 to 100% they see as sufficient to:

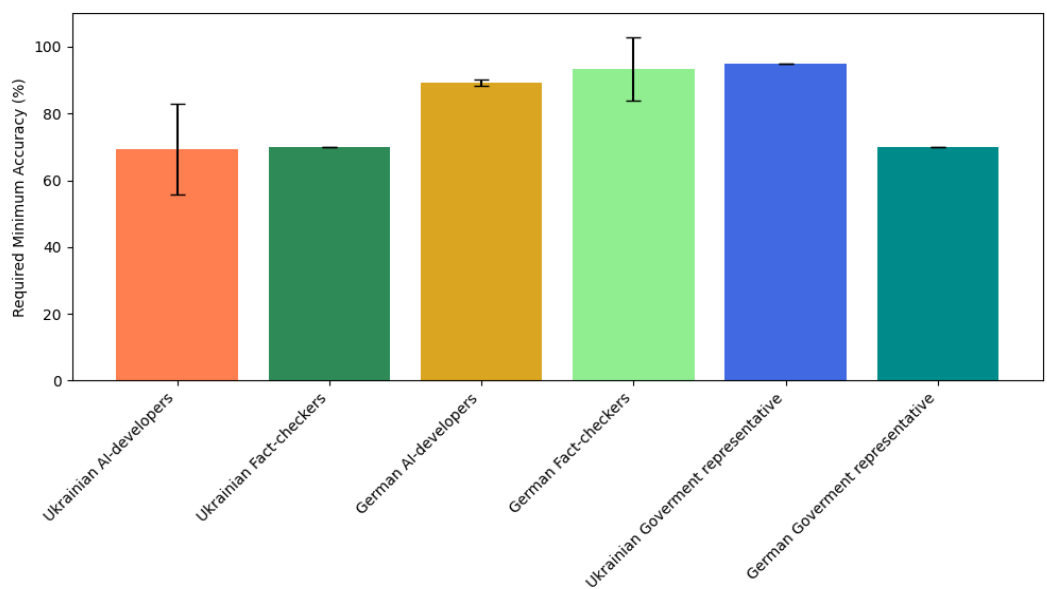

*Figure 1*. *The results of the quantitative survey on required accuracy perception for AI tools for fact-checking (n=13).*

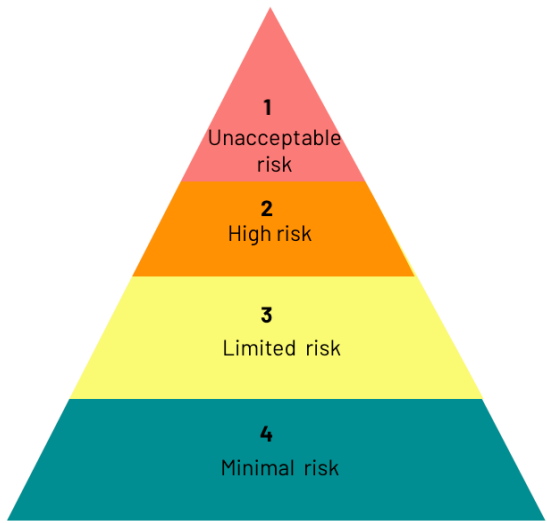

*Figure 2*: EU AI Act AI application risk categorisation.

(1) for fact-checkers - try implementing it into their fact-checking routine

(2) for AI developers - offer the application partner or client to use it

(3) for government representatives - to implement it into the governmental info-space monitoring system.

We can see that, on average, Ukrainian respondents among AI developers and fact-checkers have a similar medium at 69.33% and 70%, respectively, although Ukrainian developers have a much higher standard deviation (±13.69), meaning more variability in answers. German developers and fact-checkers also show rather harmonised understanding at a much higher medium of 89.33% and 93.33%, with fact-checkers showing more volatile answers with a standard deviation of ±9.42.

German satisfactory accuracy estimations are, hence, on average 20% stricter than Ukrainian ones. Interestingly, for the governmental responses, we see an opposite tendency with 70% from the

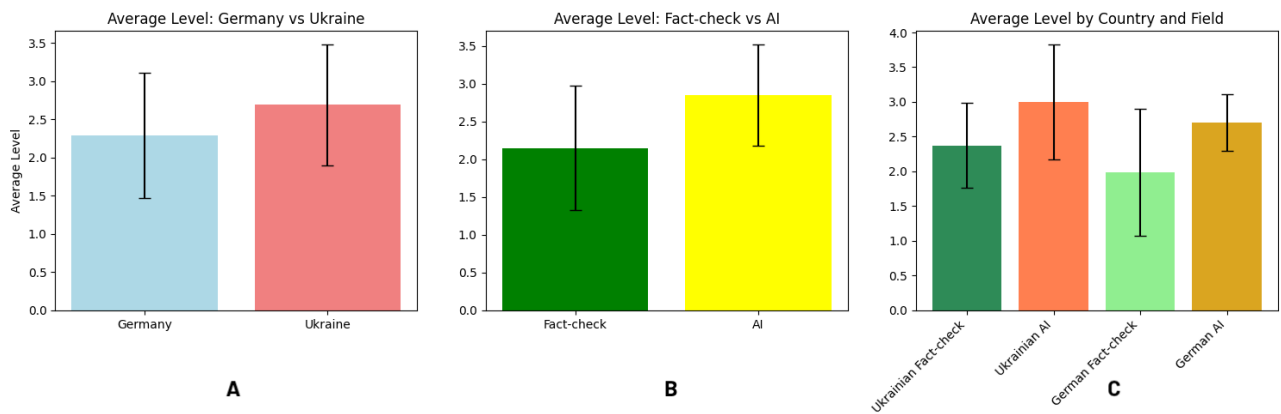

**Figure 3:** *The average level of AI fact-checking tools acceptance aggregated per country and specialisation. Based on the risk categorisations lower bars closer to 1 indicate higher risk attributed, while higher bars closer to 4, lower risk. n=11*

German government and 95% from the Ukrainian one. However, we only have one response for each of the governmental functionaries, and thus the results might not be representative.

In addition to the interviews, based on the existing literature, we created a ranking form with 16 existing and potential AI applications for fact-checking and asked our respondents to rank it based on the EU AI Act risk categorisation scale:

1. Unacceptable risk applications involve, e.g. social scoring, facial recognition, and manipulations and are banned in the EU.
2. High-risk AI applications include AI applications used in the transportation system, education, employment processes, the justice are heavily regulated and require human oversight.
3. Limited risk AI applications are chatbots, and emotion recognition systems, and have mandatory transparency requirements, specifically that the users should be warned that they deal with AI actors.

4. Minimal risk applications include e.g. AI in video games, spam filters, and spell checkers; which are not subject to specific regulatory obligations. However, providers of these systems are encouraged to voluntarily adhere to codes of conduct that promote the responsible development and deployment of AI technologies.

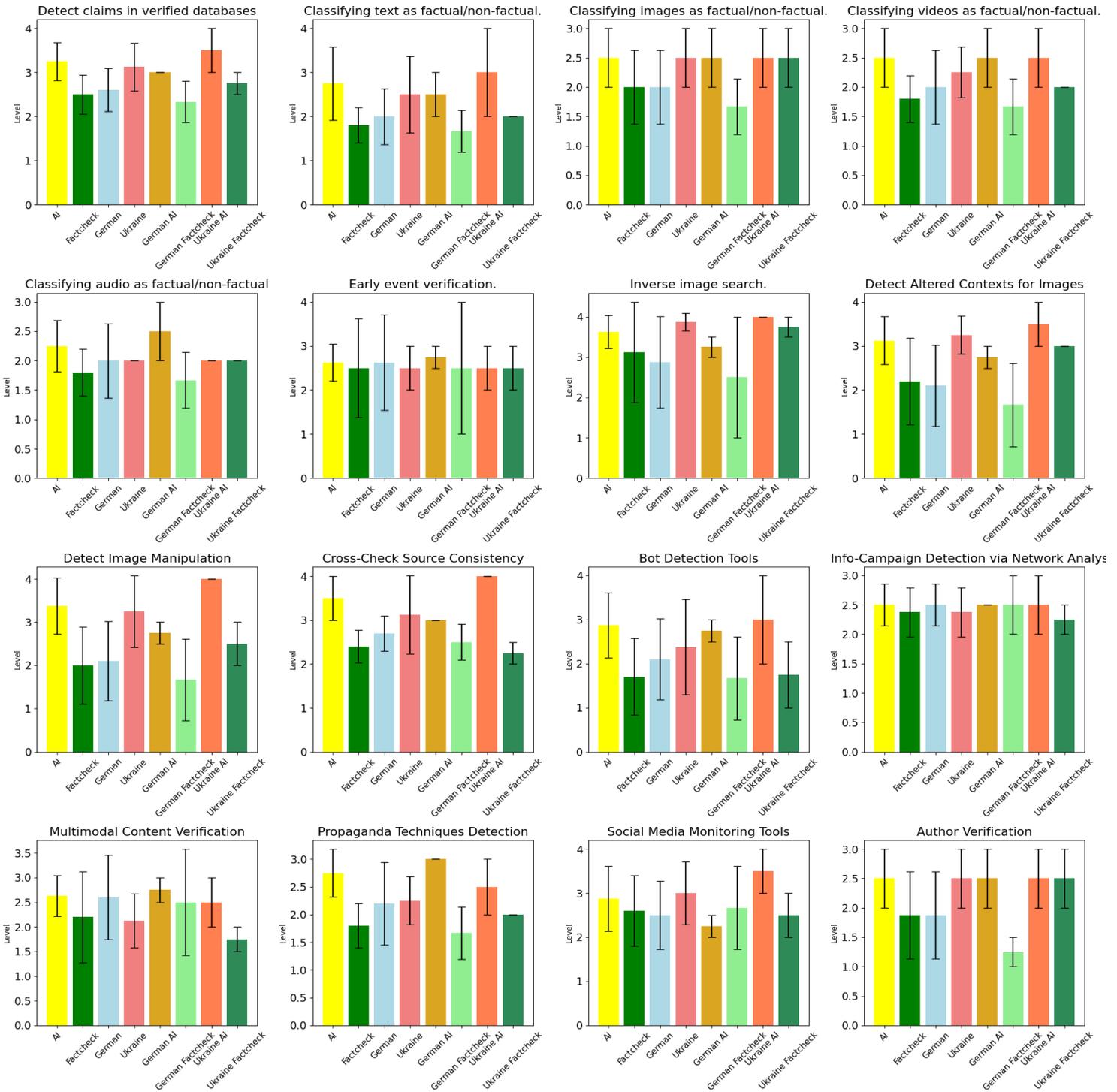

*Figure 4:* *Risk level categorisation of various AI tools for fact-checking.*

In Figure 3(A) we can see that Ukrainian respondents (both developers and fact-checkers) on average attribute lower risk categories to fact-checking tools with a 2.7 on average in comparison to 2.3 in the case of aggregated German respondents. Both are, however, positioning AI fact-checking tools on average between the High and Limited risk categories, both requiring regulation, transparency and human oversight.

We can also see a similar tendency between aggregated fact-checkers and AI developers of both countries. On average fact-checkers are more cautious with ranking these tools (average = 2.2), never attributing a score lower than 3 (Limited risk), while AI developers on average attribute 2.8, with standard deviation indicating the presence of category 4 (Minimal risk) in the answers (see Figure 3(B))

Overall, Figure 3(C) illustrates how fact-checkers of each country have lower acceptance of the tool in comparison to same-country AI developers, with an average of 2.4/3.0 (fact-checkers/AI developers), and 0.6 points difference for Ukraine and 2.0 /2.7 (fact-checkers/AI developers) for Germany, with 0.7 points difference. Both German fact-checkers and AI developers' averages are lower than Ukrainian evaluations with a difference between German fact-checkers and Ukrainian developers of a full point.

The most harmonised views among all respondent subgroups are observed for Information Campaign Detection using Network analysis and early event verification with all groups averaging around 2.5 for both applications. In all other applications AI respondents attribute lower risk to the applications than their fact-checking colleges.

This study demonstrates differences in perceived risk associated with other AI applications. A high discrepancy between AI developers and fact-checkers can be seen in detecting image manipulation, classification of text and audio as factual/non-factual, bot detection, cross-checking source consistency and propaganda detection techniques.

On average, the Ukrainian AI developers group is the most optimistic about the associated risks of the tools overall and in comparison to their German counterparts. Ukrainian fact-checkers are, however often more sceptical than German ones. This can be observed for info campaign detection with social networks, cross-checking source consistency, multimodal content verification, and social media monitoring tools. Nonetheless, German fact-checkers are the most sceptical group. They are considering equally risky classifying text, images, videos and audio as factual/non-factual, all averaging around 1.7. They are also sceptical about the detection of altered context for images, bot detection, author verification, propaganda techniques detection, and detecting image manipulation (1.5-1.7).

In some applications, German respondents have a more lenient attitude towards the associated risks. These include multimodal verification (2.5 to 2.0 Ukrainian average) and early event verification (2.5 to 2.4 Ukrainian average), and the effect can be observed across both fact-checkers and AI developers. They also have on average similar views on audio deepfake detection (around 2.0).

The least risk on average was attributed to Image Reverse Search (3.5) and the detection of similar fact-checked claims to the new query, both already widely used or developed applications in fact-checking processes. Other applications would range between 2.0 and 2.5 on average.

It is important to highlight that our sample size is insufficient to make conclusions about these effects and correlations. A notable disparity can be seen in the AI developers group, where both Ukrainian developers are working in a start-up at non-managerial positions, while both AI representatives from Germany are from academic settings, are from the same university, and are professors and junior group leaders. However, most of our representatives represent the opinion of their organisation, and thus, we can treat them as an aggregated average of other group members.

# Policy analysis

## Most Urgent Tools to Develop

German and Ukrainian fact-checkers agree on the need for advanced AI tools, though their specific priorities reflect distinct concerns and challenges faced in each country's media environment.

Many suggested that a tool capable of scanning large volumes of documents early in the investigative process to trace the origins of specific citations or claims would save significant time, even if human verification remains essential. At CORRECTIV, they envision tools that can narrow down information, such as sorting through 20 studies or filtering hundreds of claims based on specific criteria.

Multiple respondents agree that future tool development depends heavily on the trajectory of disinformation. For instance, while current deepfake technology is not widely challenging yet, they anticipate a need for more robust detection tools as it advances. Audio manipulation detection, another less-explored area, also may emerge as a priority. Distinguishing legitimate edits from fake audio manipulation could aid journalists, given that audio, unlike images, does not always "look strange" at a glance and may be more readily trusted by audiences. Real audio from radio program snippets can also be edited and smoothed, triggering such systems.

Another task in this field, which does not seem to be addressed well yet, is voice similarity models. Such a model should be able to say with high confidence if speech recordings belong or do not belong to the same person. A crucial aspect is also robustness to generate imitations of real people.

As of now most famous cases include Biden's robocall[5] and several mayors of European cities duped into deepfake video calls with the mayor of Kyiv[6].

Deutsche Welle representative also underlined that continuous work is needed to effectively detect "cheap fakes" that, though rudimentary, remain effective in spreading mis- and disinformation. DW advocates for making these tools accessible to a wide audience, such as integrating them into social media platforms, offering browser plugins, and creating user-friendly apps that cater to everyday users like students and the elderly.

In Ukraine, fact-checkers at VoxUkraine prioritize tools that enhance source quality and reduce issues like AI "hallucinations" and cherry-picking. They seek improvements in AI's handling of screenshots and want a reduction in false negatives to make AI-driven fact-checking more reliable. As media monitoring becomes increasingly vital, both German RBB and Ukrainian Detector Media see potential in creating tools that alert users to emerging trends across multiple social media platforms and identify disinformation campaigns and deepfake content as it evolves. CORRECTIV also seeks better data monitoring and trend analysis tools, which would allow fact-checkers to go beyond verifying individual claims to examine broader narratives and how specific claims feed into these larger narratives.

---

[5] https://www.theguardian.com/technology/article/2024/aug/22/fake-biden-robocalls-fine-lingo-telecom

[6] https://www.theguardian.com/world/2022/jun/25/european-leaders-deepfake-video-calls-mayor-of-kyiv-vitali-klitschko

Additionally, in Detector Media, they emphasize the need for network analysis tools that could map influence networks and track citation patterns across social media platforms like Telegram, enhancing the ability to trace disinformation. Sophisticated cross-platform scraping tools, capable of gathering data from both Telegram and Facebook, would allow for a more comprehensive approach to monitoring, addressing the challenges of fragmented information spaces.

The need for cross-platform analysis is echoed by AI developers and researchers, who face growing restrictions as social media platforms limit academic access to new data. Platforms such as Reddit restrict access entirely, while others like Meta, TikTok, and YouTube implement lengthy application processes or charge high fees, restricting real-time data access and complicating reproducibility and open science efforts essential to academia. Currently, only a few platforms, like Telegram, allow relatively unrestricted data collection, but this alone is insufficient due to Telegram's limited demographic reach.

Importantly, Detector Media underscores the irreplaceable role of human judgment, especially in understanding context. Disinformation is not merely an informational challenge but a component of the broader "hybrid war," where political, economic, and social factors intersect. Consequently, while AI tools are instrumental in building resilience against disinformation, they remain one part of a complex ecosystem that still relies heavily on human expertise and situational awareness.

Our respondents from the AI development agree on the need for real-time fact-checking and multimedia verification in terms of manipulated and generated content as the most pressing needs in today's information landscape. Additionally, there is a need for explainable AI (XAI) tools specifically designed for fact-checking. These tools would offer transparent explanations of how conclusions are reached, building trust and supporting informed judgments.

Both News-Polygraph and noFake stress the importance of tools for tracing the origin and spread of disinformation across social media platforms. Such tools would allow fact-checkers to map the progression of narratives, identify key amplifiers, and gain insight into how disinformation campaigns shape public opinion—especially critical before elections when disinformation is often at its peak. In Ukraine, Mantis Analytics points out the value of end-to-end systems capable of verifying news by checking sources, tracking first appearances, and assessing content credibility. While simpler prescreening tools offer preliminary checks and broader accessibility, more advanced verification systems could provide deep, context-specific analyses, particularly useful for fact-checking professionals.

## Criteria for AI Tool Testing

Integrating AI into fact-checking workflows requires transparency, accuracy, and explainability while preserving user autonomy and public trust.

Detector Media emphasizes seamless integration into team workflows and workspaces, ensuring that AI tools complement rather than disrupt established processes. VoxUkraine highlights the need for tools that can manage varying levels of complexity, handling both straightforward and more challenging cases. For instance, if an AI tool is tasked with identifying primary sources, it must search in a cross-lingual way and prioritize credible media, not merely the most readily available sources. The goal is to ensure that the AI can distinguish between high-quality sources—such as official bank statements over secondary reports in local news and tertiary sources from lower-quality social media news—to improve information reliability.

German fact-checkers place a strong emphasis on accuracy, transparency, and reliability, with specific accuracy requirements that vary based on the context of each project. Accuracy is critical, but credibility goes beyond just a percentage figure; even a minor inaccuracy (like a 0.5% error) can impact public trust and damage reputation, which is hard-earned and continuously challenged in the current political climate. Therefore, transparency in AI processes is essential; the AI tool should allow users to trace back the decision-making process to understand how conclusions were reached. This level of transparency is key to maintaining public trust in fact-checking institutions.

For some projects, accuracy requirements can be lower (around 80%) without compromising effectiveness, such as in disinformation monitoring, where even partial identification of misleading content is beneficial. However, higher accuracy is necessary in cases like audio manipulation detection, where outputs must be consistently reliable and explainable to ensure journalists can make informed assessments. RBB also values reproducibility, as fact-checking requires consistent answers to maintain reliability—a challenge for tools like ChatGPT that can generate varying outputs with each query. Deutsche Welle (DW) evaluates AI tools based on performance consistency across real and fake classifications. They found that True Media, with its balanced performance for both real and fake content, met their criteria, whereas other tools like Hugging Face and Hive fell short.

The role of **explainability** and **transparency** varies depending on the specific task, with flexibility in AI-generated explanations highlighted by all German media representatives. RBB notes that while detailed explanations may be essential in high-stakes scenarios, they are less necessary in simpler tasks like Telegram monitoring, where journalists primarily need quick information without the underlying technical details. A score indicating confidence would suffice in many cases, though RBB appreciates the option for a drop-down explanation feature, especially when the model's confidence level is uncertain or when journalists want insight into the AI's decision-making process. DW echoes the sentiment, noting that not many tools currently offer this level of optional explainability, but that it would be highly valuable if available. Importantly, DW emphasizes giving journalists the ability to verify information independently before consulting AI tools, minimizing bias and ensuring that AI remains a supportive tool rather than a primary source.

CORRECTIV stresses that in tasks requiring precise article verification, traceability and explanation are essential. However, for simpler tasks that primarily involve narrowing down large data sets, transparency is less critical. Nonetheless, they believe those applying the tools should have a good foundational understanding of the principles behind them to use the results responsibly.

In Ukraine, VoxUkraine sees explainability as especially useful in handling visual media, where pixel-level indicators can save time when verifying images or videos. Having a detailed path that shows source origins and spread points is also highly valuable, as it simplifies the process of explaining the findings to the public. On the other hand, Detector Media prefers minimal AI-generated explanations, noting that they would prefer to conduct their own analysis and interpret the findings independently in any case.

# The promise of Scaling

Towards AI efficiency and adaptability while enhancing collaboration and human AI literacy.

Scaling fact-checking with AI requires achieving a complicated balance between speed, accuracy, accessibility, and adaptability. Fact-checking organisations in both Ukraine and Germany recognise that AI can accelerate their workflows, particularly in data-intensive tasks such as document analysis, early disinformation detection, social media monitoring to identify checkworthy claims and events, and multilingual content processing. AI might one day generate search keywords autonomously, which could streamline processes currently managed by humans. However, DW also experienced that social media monitoring systems can be skewed by bots and trolls, as disinformation actors are aware of these tools and adapt accordingly.

Similarly, RBB sees potential in AI for handling large data volumes, including audio and deepfake detection. They note that while AI may not be perfect, it provides an additional layer of oversight, helping them check content more systematically and address fatigue-induced errors common in human-only workflows. CORRECTIV supports AI's role in scaling fact-checking efforts but distinguishes between two key functions: identifying false claims across multiple platforms and helping with editorial preparation of fact-check text for publishing, where AI can help create accessible content summaries, e.g. a simplified version, or versions for diverse audiences, shorter versions, or video scripts that can broaden fact-checking reach.

Ukrainian organizations such as VoxUkraine and Detector Media focus on AI's potential to contextualize disinformation by providing background on evolving narratives, especially as disinformation patterns often recur globally. AI could become invaluable for tracking narrative patterns across regions and offering context-sensitive information to counter emerging disinformation. Especially, AI could be a useful guide when local analysts and fact-checkers lack context-specific knowledge and a culturally aware understanding of other regions to process world news correctly.

From the perspective of AI developers, scaling up fact-checking remains challenging due to issues like data diversity, model robustness, and computational resources. The existent variability of disinformation across languages and cultures requires diverse, up-to-date training data for the models to remain effective. The rapid evolution of disinformation tactics further complicates this, as models must frequently adapt to new manipulation techniques, requiring considerable computational resources. Moreover, curated, updatable benchmarks with unbiased, high-quality data and agreed-upon labelling schemes are not systematic, so consistent testing of the models and their comparison is problematic in this field.

In the age of the Large Language Models, in fact, smaller models fine-tuned and tailored for specific tasks are still widely used and most economically viable, as they provide faster, cheaper, and more precise responses, making AI-driven fact-checking and research in this field more accessible to organisations with limited resources. Non-coincidentally, the current cutting-edge research in Natural Language Processing focuses on making Large Language Models smaller without losing efficiency. In this light, the current GPT-4o model (OpenAI, 2024), Mamba (Gu & Dao, 2024), and smaller NVIDIA (e.g. optimised Llama 7b fine-tune[7]) solutions are also steps in the direction of cheaper and more efficient fact-checking. The resources spent on AI research for mis- and disinformation detection are very scattered, which makes it difficult to find joint approaches tackling mis-and disinformation. E.g., in the EU there are many research projects concerning mis- and disinformation detection, and they barely communicate with each other or find ways to spend the given resources effectively. Clearly, there is a need to foster more communication within the funding clusters and among different similar clusters.

Finally, most of our respondent fact-checkers and developers underlined that while many AI tools are already on the market and available free of charge, raising human skill and awareness of these tools, addressing their fears is also essential for scaling fact-checking, highlighting the need to train analysts and journalists in current AI tools and methodologies.

---

[7] https://catalog.ngc.nvidia.com/orgs/nvidia/teams/nemo/models/llama-2-7b

While several of our German respondents constantly organise various training lectures, and workshops for different target audiences to increase media literacy and explain fact-checking basics, Ukrainian fact-checkers and AI developers underline that there is not enough such initiative nor funding for this purpose in Ukraine.

## Reducing Skepticism Toward AI in Fact-Checking

Training and Awareness are key to combating AI scepticism

To reduce scepticism around AI in fact-checking, experts across Ukraine and Germany agree on the importance of training, transparency, and fostering user awareness. Fact-checkers like RBB emphasize viewing AI as an additional tool rather than a replacement, dispelling fears that AI might replace human roles entirely. Deutsche Welle (DW) acknowledges the negative public perception AI has acquired, especially with the rise of synthetic media like deepfake images and videos. DW regularly trains students and journalists to bridge the "AI awareness gap" and to reframe AI as a helpful tool with diverse, constructive applications. CORRECTIV underscores the need not just for AI acceptance, but for a deeper understanding of how it works. They suggest that every media organization could benefit from small, hands-on AI hubs where employees could actively experiment with AI tools, fostering a culture of collaborative digital innovation within media houses.

In Ukraine, Detector Media highlights the broader role of AI in countering Russian disinformation, noting that AI's potential is linked to political commitment and coordinated efforts across NATO countries. For Ukrainians, awareness of AI's role in addressing hybrid warfare—covering economic, energy, historical, and political manipulation—is critical. VoxUkraine suggests that normalizing the creation of fact-checking departments within media organisations could reduce disinformation from the start, envisioning dedicated fact-checking departments to ensure accuracy before publication.

From the perspective of AI developers, increasing AI acceptance should be focused on transparency and user control. News-Polygraph and noFake project advocate for explainable AI (XAI) features that allow journalists to trace AI decision-making paths, understand the evidence behind flagged claims, and adjust settings to align with journalistic standards. Such transparency fosters trust by allowing journalists to see and, if needed, override AI decisions. A feedback loop that incorporates journalists' expertise into the AI's evolution further enhances acceptance, establishing a co-adaptive model where AI becomes a reliable assistant.  Ultimately, they believe that high accuracy and reliable performance on novel data will be essential to build lasting trust; when AI tools consistently demonstrate their value, they will be readily accepted.

Ukrainian developers, in contrast, underline that AI adoption is partly driven by social validation; as more users adopt AI, its acceptance and usage will naturally grow. Moreover, it will naturally create more feedback for the developers to improve the models. This, in its order, will add to the positive loop, increasing AI acceptance.

# The role of the regulation

Regulation of Online Media and Public Broadcasting in Ukraine

In Ukraine the Media Law 2849-IX[8] was registered on the 2nd of June, 2020, approbated on the 13 of December, 2023 and formally came into force on the 31st of March, 2023.

Our Ukrainian media respondents expressed scepticism in terms of this law having real consequences for the Ukrainian media landscape. Especially the responsibility of the media for spreading intentional disinformation and the methods of its enforcement is still vague. One of the suggestions included the creation of an independent journalist ethics and standards commission to review such cases and well-thought-out guidelines to be established. In particular, these should include a classification of the seriousness of the false claims based on the potential consequences and whether the case can be classified as accidental misinformation or intentional disinformation[9].

Now, that such cases do not receive a proper reaction, the spreaders realise their impunity.

Many news outlets could benefit from structural improvements, such as establishing well-defined editorial boards and implementing a clear hierarchy of responsibility to strengthen quality assurance practices and internal control mechanism guidelines. These, according to the Ukrainian representative, may not necessarily copy rather excessively rigid German examples. Considering the 6-eye principle, for instance, with new team members joining more people are required to review the materials.

The problem with such controlling bodies in the current Ukrainian system in a young democracy without well-established and functional institutions is that they can be manipulated by the political parties. In this sense, the German example of media regulators being regional and not country-level actors, independent from the government but still with their own enforcement mechanisms, may serve as a promising template.

However, subletting the responsibility for establishing AI regulations for journalism in the Ukrainian political landscape and unitary structure of governance, seem to make even less sense than in federal Germany. Clearly, to provide the same level playing field for media from different regions, such regulation should be shared and come on a state level.

The relative lack of established media regulations in Ukraine offers a unique opportunity to build a media framework that incorporates successful practices from Western media models while deliberately avoiding their notable drawbacks. One such example is the principle of **balance** or **impartiality**, where equal coverage is traditionally given to both sides of a conflict.

In today's post-truth information landscape, however, strict adherence to this principle can lead to **false balance** (or **false equivalence**), which arises when unequal sides are treated as though they are equally credible or substantiated. This issue is familiar to the Ukrainian public, who have seen respected outlets like Reuters give equal weight to Russian and Ukrainian perspectives on the war. As the Head of the VoxUkraine fact-checking team points out, this approach is not only ethically questionable—since it equates the victim with the aggressor—but it also risks irrevocably shaping Western public opinion. Even if the Russian narrative is later discredited and the article edited, Western readers, lacking context, may already have formed a perspective that frames the conflict as ambiguous or "not entirely clear-cut." The adaptation of the media regulations to encompass "alternative facts", half-truths as well as conspiracy theories and many other modern phenomena, and the way how this will work with freedom of speech as a fundamental right, will clearly be a hard thought battle, the success and failure of which is instrumental and vital to the future of democracies. Ukraine has the potential to lead by example, pioneering not only effective principles for addressing these issues but also setting forward-thinking standards in AI regulation for media.

---

[8] https://zakon.rada.gov.ua/laws/show/2849-20?lang=en#Text

[9] From here forward, we always define misinformation and unintentional spread, and disinformation as intentional.

**A. PERSPECTIVES FROM THE NATIONAL SECURITY AND DEFENCE COUNCIL OF UKRAINE**

*Based on the interview with the Head of Communications of the National Coordination Centre for Cybersecurity at the National Security and Defence Council of Ukraine, Anastasia Ostrovska.*

The **National Security and Defence Council of Ukraine (NSDC)** coordinates and implements Ukraine's national security and defence policies, focusing on areas such as military strategy, cybersecurity, and countering disinformation. It acts as a crisis management body, advising the President, aligning various agencies, and fostering international partnerships to enhance Ukraine's resilience against internal and external threats.

In NSDC, AI's role in formal communication, such as drafting letters to international partners, is acknowledged, although there's reluctance around critical information sharing with AI, especially among older generation officials who are less familiar with AI's benefits and risks.

They also underline the need for better filtering and context-based analysis in the context of info-space analysis for platforms like Telegram, which is highly monitored in Ukraine but still hosts harmful content (in addition to disinformation e.g., phishing, and drug trade). In contrast, collaborations with Meta have been more effective due to Meta's robust fact-checking.

To increase trust and adoption of AI-based information monitoring and fact-checking tools within government, transparency and accountability are essential. Although AI tools are already assisting with preliminary analyses and insights, officials still rely on primary sources for final validation. One proposal suggests developing a centralized, AI-enhanced "Wikipedia-like" knowledge base specifically for governmental use. Such a repository could be backed by NATO, the EU, or other international bodies, allowing trusted contributors to input verified information and ensuring that AI-driven data remains secure and up-to-date. In this vision, private companies, security experts, and governmental bodies would contribute collaboratively, enhancing resilience against disinformation. Additional security features, like a "bug bounty" program, could proactively identify and resolve system vulnerabilities, demonstrating robustness against potential interference from hostile actors like Russia.

**B. PERSPECTIVES FROM THE FORMER MINISTRY OF INFORMATION POLICY OF UKRAINE**

*Based on the interview with former Deputy Minister of Information Policy, Dmytro Zolotuhin.*

The Ministry of Information Policy was established on December 2, 2014, to oversee Ukraine's information policy and counteract disinformation. It was dissolved on August 29, 2019, during a government reorganization and in 2020 it was merged into the Ministry of Culture and Information Policy later in 2024, it was renamed the Ministry of Culture and Strategic Communications.

Ukraine's approach to disinformation resilience began with the 2017 Information Security Doctrine, emphasizing the delicate balance between monitoring harmful information and upholding free speech. This included a national media monitoring effort, to detect and respond to disinformation based on 50-60 criteria—though not yet AI-driven. Initiatives included fact-checking, media literacy, and social media disclaimers to alert citizens to potential dis- and misinformation. However, traditional fact-checking was deemed insufficient due to limited reach and lengthy implementation.

By 2018, the ministry saw automatisation and AI as crucial for enhancing detection but recognized the challenges of maintaining unbiased, nuanced outputs. In the Ukrainian context, fact-checking organisations can be manipulated. Rather than binary fact-checking results, they advocated for a probabilistic, percentage-based AI approach, allowing for context-driven disinformation responses.

Ultimately, the ministry viewed disinformation resilience as a complex field where AI could compel adversaries, like Russia, to adjust and finance new propaganda tactics—cost-effective compared to traditional methods. However, they noted that international donors often hesitated to fund such AI advancements, suggesting further development is essential for adapting to modern information warfare

---

## Regulation of Online Media and Public Broadcasting in Germany

*Based on the interview with Benedict Föll from Medienanstalt Berlin-Brandenburg (MABB) and Hanna Katarina Müller (head of the Hybrid threats unit at the German Federal Ministry of The Interior (BMI)*

**A. THE CASE OF MABB IN BERLIN-BRANDENBURG**

The Media Authority Berlin-Brandenburg (MABB) oversees online media and public broadcasting, focusing on both regulation and funding initiatives that promote information and news literacy, particularly through local journalistic projects. Funding is directed at projects that enhance public understanding of information, and MABB also regulates compliance with journalistic diligence, a legal standard that requires accuracy and responsibility in reporting. MABB's authority is independent from direct government control, ensuring that decisions on content do not reflect governmental influence and that freedom of speech is upheld.

In practice, MABB has legal avenues to address cases where journalistic diligence is compromised. For instance, if a news outlet or fact-checking website publishes false information, MABB can issue a complaint or, in severe cases, prohibit the publication of specific content or more usually specifically the part of the content which is problematic.

Recently, MABB encountered its first case involving AI-generated content on a fact-checking site, where an obviously false claim was published due to reliance on AI for editing the fact-check material. In such cases, MABB's intervention may not necessarily prohibit the full publication but rather the specific negligent sections. As of now such negligence cases only focused on the lack of rigorous verification of generated text for publication.

MABB is cautious about the exclusive use of AI tools for fact-checking, e.g. in detecting deepfakes. They stress that how claims are formulated and verified is critical, and AI verification should be done with human oversight to avoid legal repercussions and uphold journalistic standards. In terms of the potentially near future, where generators improve so much that only AI can detect AI-generated content, the media regulation rules seem as of now unprepared.

**B. PERSPECTIVES FROM GERMAN THE MINISTRY OF THE INTERIOR**

The Ministry of the Interior's division focused on hybrid threats plays a leading role in addressing disinformation. In this department, the new ministry initiative also organises a specialized office for disinformation resilience: "Central Central Office for the Detection of Foreign Information Manipulation" (ZEAM). Its efforts will be divided between detection (using AI) and resilience-building initiatives, such as funding projects to enhance media literacy through citizens' assemblies.

A dedicated task force coordinates various agencies, including the Ministry of Foreign Affairs and the Ministry of Justice, to centralize efforts against disinformation.

According to the Head of the Division "Political Systems; Hybrid Threats, Disinformation" and head of the development team of the newly established ZEAM, Hanna Katerina Müller agrees that AI could significantly enhance detection capabilities by filtering vast amounts of social media content—a task too resource-intensive for human analysts alone. Automation could help address limited human resources and offer initial insights, allowing human teams to focus on high-priority cases. Additionally, AI could assist in promoting media and technical literacy, tailoring educational approaches for diverse target groups.

The Ministry's main concerns revolve around biases, transparency, and public perception. Given AI's potential to reflect societal biases, it is essential to ensure impartiality and accuracy in AI-assisted outputs. Transparency is also a priority; but transparency of a different kind: the government must openly communicate its use of AI for disinformation detection and clarify the criteria for monitoring specific content. The Ministry notes that government experts often lack deep technical knowledge of AI, necessitating reliance on external companies and raising questions about informed oversight.

Beyond accuracy, reliability, security, and transparency are equally critical, ensuring that AI tools meet governmental standards for public accountability and data integrity. Cybersecurity is also a key consideration, as robust protections are essential to maintaining trust in AI-powered disinformation efforts.

Enhancing AI acceptance within government agencies requires both expertise and clear communication. The Ministry suggests creating roles that bridge technical knowledge and policy, as decision-makers often lack digital fluency. Establishing a common language for discussing AI risks and capabilities could foster greater trust. Furthermore, experimental flexibility is crucial; delays in adopting AI can render tools outdated, making it essential to create safe spaces for trial and error, along with adaptable financing to support ongoing AI development.

**C. EUROPEAN UNION AI ACT**

One of the obvious sources of inspiration for updated media regulations can be found in the European Union AI Act[10]. Based on the EU AI Act, the risk level depends heavily on the level of automation and the resulting actions taken. In case the applications work fully autonomously and result in blocking or deleting the content from, e.g., social media platforms, they apply to high-risk applications, and should not be used without human-in-the-loop in fact-checking.
According to the named regulation, in case the results are to some extent (depending on the thresholds of automation) being shown to human users, then the applications would mostly apply to the limited risk category.
The other existing legislative basis is the Digital Service Act (DSA). DSA is an EU regulation aimed at creating a safer, more transparent online environment by setting clear responsibilities for online platforms and digital services. It focuses on reducing the spread of illegal content, enhancing user protections, and ensuring accountability for large platforms, particularly those classified as Very Large Online Platforms (VLOPs). Key measures include content moderation standards, transparency on algorithms and advertising, risk assessments, and stronger oversight to safeguard fundamental rights.

---

[10] https://artificialintelligenceact.eu/

# Policy Recommendations

## I. New Media Regulation

### Ukraine

To be able to implement media regulations for AI usage in journalism and fact-checking, the first corresponding legal and structural basis is necessary to establish:

1. Adopt current laws about media to create an independent media regulatory body in each region, independent from the government, and thus difficult to use for censorship. A higher body of the national-level committee can also be created for more difficult cases (such as the Commission on Licensing and Supervision (ZAK)). German State Media Treaty (Medienstaatsvertrag) can be used as a starting point but adapted to the Ukrainian structure of governance and local legislative specifics.
2. The members should be employed through a transparent public process out of representatives of all major regional media for the local body and national level media for the central committee.
3. This body should have the power to regulate and treat cases of media negligence, defamation, misinformation, and disinformation. Negligence and unintentional cases of misinformation should result in an administrative processing fee and the legally binding requirement to remove part of the text. In case of non-compliance, the case should go to the court of justice. Grounded suspicions of intentional defamation and disinformation cases should be then passed to the court of justice with the collected evidence. The law for the intentional spread of disinformation should be worked through and approved by the Ukrainian Parliament. The main category to prove in the investigation process should be the intent and the third party whose interest the article promoted. The important work for the lawmakers would be to ensure that this law does take the shape of the notoriously known "foreign agent media" law, which as of 2024 exists in Russia, Belarus and Georgia, but rather drew inspiration from e.g. aforementioned German State Media Treaty, European Union Digital Services Act (DSA), Network Enforcement Act (Netzwerkdurchsetzungsgesetz or NetzDG).

### Germany

For AI regulation in fact-checking to be effective, it must be underpinned by robust and coherent disinformation regulation. In Germany, significant room for improvement remains in the context of disinformation resilience laws. Historical sensitivities stemming from the Nazi regime and World War II have left the concept of "censorship" vulnerable to manipulation by radical political parties, creating friction in reforming public consensus on Freedom of Speech in the digital era and the context of modern informational warfare.
This complexity has led to regulatory measures like blocking Russia Today (RT) from public broadcasting[11], which was justified by the absence of a broadcasting license rather than addressing RT's most dangerous function: serving as a tool for foreign disinformation campaigns and a threat to German democracy and national security. Such indirect approaches to regulation are not unique to Germany; similar patterns are evident across Western institutions. For instance, the ICC's arrest warrants against Vladimir Putin and Maria

---

[11] https://www.dw.com/en/russias-rt-channel-blocked-by-german-regulators/a-60635397

Lvova-Belova focused not on the breadth of documented war crimes since 2014 but specifically on one case: child deportations, a crime they confirmed themselves.

To foster greater public understanding and trust in governmental decisions, it is crucial for Western institutions, including Germany, to address violations directly and transparently. Laws must call infringements by their true names, rather than those easier to justify under current legislation, if we hope to enhance public awareness of the dangers these regulations aim to mitigate.

The following steps could help achieve this goal:

1. Facilitating Public Consensus: Invest and encourage public debates featuring intellectuals, conduct surveys, and together with civil society and governmental institutions collectively shape a new social contract that balances freedom of speech with the need to counter disinformation effectively.
2. Legislative Reform**:** Update and adapt national legislation to address emerging challenges in the digital era.
3. EU-Level Harmonization**:** Align national laws with broader EU regulations to ensure consistency and effectiveness across member states.

---

## Germany and Ukraine

However, legislative frameworks like the AI Act might not fully enforce transparency requirements across all actors, making detection tools crucial.
Regulation for AI usage in journalism and fact-checking should be implemented:

It can be based on the current standards adding on certain points specific to AI usage to ensure clarity, which is now absent and is interpreted by each media outlet based on current regulations. These points could lay the foundation for such new regulations:

0. Fully autonomous AI systems can only be applied for media monitoring to detect and rank a list of checkworthy material, which then should be curated by human fact-checkers.
1. In all other applications, AI has to be seen as a tool, so it does not substitute a journalist, and the 6-eye principle can be preserved.
2. Non-biasing principle: human should first make up their mind and only then use AI.
3. Concerning generated content: if only AI can detect the generated nature of the content, and the indicators it gives are non-interpretable, and in case other classical journalistic investigative approaches also have provided no solution, the verdict should be "partially verified", or "AI-verified".
4. Previously untested, or insufficiently tested tools' predictions should not be taken into account.
5. A majority vote of 3 different tools if possible about confidence of the models should be taken into account.
6. Preference should be given to the tools which explain the decision, also to know when the explanation does not make sense, and the prediction should not be considered reliable.
7. Media outlets should create experimentation and educational hubs for employees to test and play with different AI tools for journalism. Workshops for AI literacy are necessary.

# II. Governmental funding strategies

## General recommendations

1) Regular AI literacy workshops are essential for government members to increase awareness of the advantages and limits of the technology and address misunderstandings and fears related to the lack of digital literacy. One of the important points here, as suggested by our governmental respondents is to strengthen the understanding that AI is not just one technology, but multiple different approaches, with various associated risk levels;

2) Establishing roles that bridge technical expertise and policy is vital, as many decision-makers lack digital fluency. For this, a possibility is to create more special AI and Computer Science master programs accepting students with humanities backgrounds in law, political science and administration with lower technical competency threshold for acceptance, and focus on broad and general familiarisation with the field, tasks and state of art. Especially in combination with technical or interdisciplinary PhD programs, such specialists can become valuable mediators between AI research and the government and governmental functionaries within regulatory bodies;

3) Within accountability standards, governments also need experimental flexibility and a safe space for trial and error. With AI rapidly evolving and here to stay, the country must adapt accordingly.

## Funding recommendations

In addition to the extension of the projects focused on AI for fact-checking in Germany currently mostly funded via the Federal Ministry of Education and Research (BMBF) and DFG, more calls from the Ministry of the Interior and Community (BMI), Ministry of Defence, Federal Foreign Office (AA), The Bundesnachrichtendienst (BND) and other ministries could increase the output of scientific output in this crucial field. Below in Section III, we also analyse the most promising technologies to be researched.

We identified several directions for funding that need more addressing.

1) Funding projects promoting AI literacy and digital information literacy are in schools, journalism, and government in various conventional and experimental formats;

2) Funding 100% TVL-13 PhD positions for digital journalism and similar specialisations on par with their AI and Computer Science counterparts;

3) Collaborative project for consortiums including European Union candidate countries and Association partners, especially in Eastern Europe (Ukraine, Moldova, Georgia) and Balkan countries (North Macedonia, Serbia, etc). Scientists from these countries have experience with powerful foreign and internal disinformation campaigns, in the case of Ukraine, and with a combination of conventional and informational warfare, bringing unique practical knowledge and solutions. Currently, most German funding calls only allow European Union partnerships, and even when the non-EU members of the consortium are allowed, they cannot receive proper funding. However, such collaborations with funded exchange stays for German and other EU researchers in the non-EU partner countries, and vice versa would strengthen EU cooperation with its candidates and association partners, bring more light into an existing and evolving variety of disinformation campaigns and narrative strategies with regional adaptations, better identify the existing network of international and local actors, helping both German and European disinformation resilience and resilience of its candidate and associative partners informational, which in the end also benefits current EU extension plans, laying basis for better resilience of the new members and closer collaboration after European Union ascension;

4) We should foster more communication within the funding clusters and among different similar clusters on national and European Union levels;

5) Projects creating updatable sharable datasets benchmarks of disinformation and fact-checks. It is essential to fund more projects focusing on benchmark creation. In recent years only several big European projects on data collection emerged (EUvsDisinfo, FERMI (Fake nEws Risk MItigatorv), the European Union's Horizon 2020 research and innovation programme under grant agreement No 825297, "Co-Inform: Co-Creating Misinformation-Resilient Societies." Benchmarks for fact-checking tools need reference points similar to educational benchmarks, which would standardize AI models against human fact-checking performance. Specifically, further improvements may be problematic without more data access to the **social media** and media outlet data, including both current and historical archives for researchers and fact-checkers.

    Some initiatives in this regard have been initiated within the policies like The Digital Services Act (DSA)[12] published by the European Commission. Under Article 40 of the DSA, vetted researchers must be granted access to data from very large online platforms and search engines. This provision aims to facilitate the study of systemic risks and the assessment of platforms' risk mitigation measures within the EU. The European Commission has initiated public consultations to define the procedures and safeguards for this data access, emphasizing the importance of transparency and accountability in the digital space. The DSA also mandates the creation of a Transparency Database where online platforms must report their content moderation decisions. This initiative seeks to provide insights into how platforms handle illegal content, disinformation, and other online risks, thereby promoting greater openness and accountability.

    However, currently, the platforms can still prevent easy research access through long transparent and over-complicated processes, and often only granting access to historical data, and under limited quota, gatekeeping the progress in this way. The current formulation regarding application for access to specific datasets in this sense plays into this problem. Moreover, the current research API agreements, for instance, with YouTube and Meta, presuppose that all data should be then deleted after the end of the project and cannot be shared with 3rd parties, which is, however, an obligatory prerequisite for the reproducibility on most of the significant AI venues and contradicts Open Science initiatives, encouraged by the European Union in parallel. The current regulatory approach, which aims to balance research needs with user privacy and data protection concerns and avoid the risks associated with uncontrolled scraping, still does not solve the problem. In the current status quo, the researchers working with social media phenomena, such as AI fact-checking, are not only disadvantaged in terms of publishing the potential results and having access to other research project results but also simply lack enough cross-platform up-to-date data, which only ensures that fact-checking and content moderation tools always lag behind and never represent the full picture.

---

[12] https://commission.europa.eu/strategy-and-policy/priorities-2019-2024/europe-fit-digital-age/digital-services-act_en

6) With the absence of data being the main research bottleneck, a possible solution could involve the creation of a **special working group consortium** with representatives of big social media, media outlets, researchers and regulators. It could be funded via the BMBF initiative and develop working solutions and infrastructure, with the end goal of the creation of the European Association for Social Media Research with the authority to provide vetted membership, as well as perform a supervisory and regulatory function, ensuring transparency, data privacy and ethical correspondence of the projects in this research field. This membership would be a pre-requisite to receive easier access to both historical and real-time data from all big social media platforms regulated under DSA. Moreover, the data could be shared among the members and also be gathered into the community dataset. The data in such cases would still obviously need anonymisation. The Association committee would still review and approve each new project needed to stay in the association and revoke membership in case of gross negligence and unethical usage. Ukrainian and other non-EU researchers, in this case, could gain membership within the European Union Candidature status or as members of EU-funded projects and consortiums.

This kind of controlled deregulation for research purposes, with high levels of transparency, would show a unique example of the EU and social media platforms for the first time coming together and finding a real actionable solution to the disinformation problem, which will provide the most efficient equal-opportunity framework for accessing, sharing, and analysing data and testing AI solutions. The organisation could also create and provide shared computing infrastructure, as the cutting-edge AI solutions are currently computationally intensive, which prevents European researchers from competing with North American and Asian research and Industry. Shared computing cluster access could also provide a more environmentally friendly solution for processing and storage.

It would also bring a significant amount of projects and interest from international collaborators from all around the globe to European Union institutions and consortiums, bringing more international funding, exchange programs and cooperation opportunities for the democratic world.

An encouragement factor for the social media platforms could be, on the one hand, the possibility of outsourcing to a certain degree fact-checking, also having access to the research solutions, which are actually tailored for the language and format specifics of their platform, while currently, this is not the case.

Additionally, a mutually beneficial solution could incorporate a European post-grad school with a thorough selection process for PhD students working on projects between the social media industry and research on the topic of automated content moderation and automated fact-checking.

These measures can improve social media content moderation transparency, and increase users' trust, decrease the number of missed and miss-moderated cases that, for example, in Germany can lead to administrative fines of up to €50 million, according to the Network Enforcement Act (NetzDG), resulting in huge economic burden.

Regulatory changes mandating data access would be essential to improve information space monitoring, making such tools truly effective across the board.

# III. AI Development standards

## Evaluation and development process for AI Fact-Checking tools

Clearly, accuracy requirements for fact-checking tools depend heavily on context, task specificity, and the stakes involved. For most classification and information retrieval applications, it is important to highlight the equal importance of precision and recall. High recall is critical to minimize missed disinformation, while high precision helps avoid false positives that could undermine user trust. The usual harmonic average between recall and precision of the F1 measure should be taken into account, but only as an additional metric. Other metrics which can be included in the process are MCC, Auroc, and ECE.

In high-stakes cases, such as investigations involving sensitive information, a recall rate of approximately 95% can be necessary to ensure minimal oversight. Beyond raw accuracy, these models must maintain reliable performance across diverse disinformation types and media formats, handling complex, context-dependent cases with consistency.

However, often as long as a model performs at a level near or better than that of human analysts, it could significantly support and scale human fact-checking efforts, compensating for factors like human fatigue and focus.

The testing process should include the following 6 stages:

1. The usual cross-validation testing, to ensure the model's performance on most relevant cases.

2. A calibration study to measure AI's confidence in its own decisions—is just as important as accuracy, as we should ensure that AI outputs are both accurate and reliable. This alignment between confidence and accuracy is key for ensuring the tool truly benefits the fact-checking process.

3. Testing on existing benchmarks, which should include most recent and historical out-of-distribution data, various genres applicable to the case (news outlets, social media), - data of different political bias, formal and informal register, different platforms in case of the social media, where also both post and comment data types might be relevant.

4. To address the topic and regional biases, one should apply data audits, upsampling and downsampling, and counterfactual analysis to maintain neutrality. By modifying inputs like names or locations, one can detect unintended biases, while adversarial debiasing techniques and further fairness tests should also be applied when applicable.

5. Explainable AI studies, using various attributional and model-agnostic techniques to make sure that the features important for the model make sense to the fact-checkers.

6. User tests - providing the model to the customer or focus group and through an iterative process of constant feedback improve the model using the real-life new examples. This not only means retraining the model on new data, but also reflecting on potential architectural changes, and changes in the loss function, in a cost-sensitive way, so that the weights of different errors more or less significant for the task at hand are adjusted. This may mean longer funding cycles for such projects.

To successfully pass such a rigorous testing process the model should be trained on high-quality data. Without it even when the metrics on the in-distribution data may be satisfactory, they become less meaningful; performance often shifts when models encounter new data, revealing limitations in the initial metrics.

For this, to be possible both benchmarks and training datasets need to be created, as discussed in Section *Funding recommendations*.

---

Promising technologies

Essentially, in the described testing process the development should loop over the 5 stages. A technology with a lot of potential in this sense becomes reinforcement learning, where the training and testing with real customers form a loop with the model being constantly updated based on the user feedback.

Smaller, specialized LLMs may be more suitable for niche fact-checking tasks than the current GPT family, as large models like GPT-4 or Gemini, while useful for initial prototypes, are often financially impractical, have a high carbon footprint in both training and deployment and limit real-time processing capabilities in media monitoring applications. Furthermore, these large models typically require extensive examples to adapt effectively to specific contexts, making them less agile and scalable for the evolving needs of fact-checking workflows. Luckily, that is the general dynamic of Natural Language Processing and the new non-transformer solutions like Mamba and smaller models produced by NVIDIA seem to go in that direction.

Anthropic's AI 2024 Claude model[13] introduces a feature called "computer use," enabling the AI to interact with a computer like a human. This includes actions such as moving the cursor, clicking buttons, typing text, and browsing the internet. This capability allows Claude to automate tasks like web searches, form filling, and other routine computer operations. While this feature is currently in public beta and primarily aimed at developers, in the correct setting it has a lot of promise for the automatisation of fact-checkers tedious research tasks, but raises clear concerns in terms of human control and oversight.

NoFake advocates for combining curated ground-truth data with explicit knowledge sources like knowledge graphs and exploring retrieval-augmented generation (RAG) techniques to improve content-based alignment in decision-making, though these methods are still in the early stages. In addition, the news-polygraph project points out the lack of standardized evaluation frameworks for assessing the correctness and meaningfulness of XAI explanations pointing to a gap in research that must be addressed to ensure high-quality, reliable explanations. They also highlight the need for AI tools that can analyze multiple content modalities simultaneously, such as combining lip-sync detection with natural language processing of audio. This layered approach would help identify inconsistencies across visual and audio cues, offering a more reliable indication of manipulation. There's limited support for detecting manipulated audio, which poses significant risks, especially in social engineering contexts, so more projects should focus on language-agnostic manipulated audio detection tools.

Computational efficiency and scalability are also priorities; with rapid information cycles and high volumes of user-generated content, deepfake detection tools must operate in real-time. Optimized model architectures and compression techniques can help meet this demand, ideally supported by centralized or decentralized computing centres that would make these tools more accessible and provide user interfaces optimised for usability. NoFake suggests an alternative approach: instead of focusing solely on detection, watermarking and provenance labels for authentic content could create a reliable standard, allowing journalism to avoid the ongoing "cat-and-mouse" dynamic between detection tools and constantly improving generation methods.

---

[13] https://www.anthropic.com/news/3-5-models-and-computer-use

Together, these advancements would equip journalists with more reliable, adaptable, and explainable tools to meet the evolving challenges of verifying digital content.

## Improving AI Tools to Meet Fact-Checker Expectations

To align with traditional fact-checkers expectations for reliability and transparency, AI tools must prioritize explainability, user control, and continuous adaptation. News-Polygraph emphasizes that AI tools should incorporate explainable AI (XAI) techniques, such as highlighting evidence sources and reasoning paths, allowing fact-checkers to follow the AI's logic. Natural language summaries of how conclusions were reached, along with cited sources, can build trust by making AI outputs accessible and understandable for fact-checkers. For reliability, consistent performance across varied datasets and content formats—text, images, and multimedia—is essential, with continuous retraining on balanced data to adapt to evolving disinformation tactics and prevent embedded biases. NoFake advocates for transparency in design, suggesting model-agnostic techniques like LIME or SHAP for generating explanations, or embedding explainability into the model architecture itself. In Mantis Analytics, the developers also highlight the possibility of using inherently transparent algorithms when possible.

Feedback loops and iterative improvements are essential for building trust. Engaging fact-checkers in the tool's refinement process allows them to identify and address issues, creating a sense of ownership and participation in its improvement, making the AI more responsive to real-world needs. Ultimately, AI tools should function as intelligent decision-support systems rather than replacements, with the level of automation adjusted to match user expertise. Defining the optimal balance between AI-driven automation and human oversight remains an area for ongoing research, critical to meeting the nuanced requirements of the fact-checking community.

Finally, developing interfaces that balance technical insight with accessibility, and help journalists understand and control AI outputs without over-relying on them, is a priority. Advising AI should also be optional and go after the initial decision of the fact-checker.

Based on our qualitative study analysis, the AI applications to prioritize should focus on Information Campaign Detection using Network Analysis and Early Event Verification. These applications:

1. Demonstrate Harmonized Perceptions of Risk: All respondent subgroups, regardless of country or profession (fact-checkers vs. AI developers), evaluated these applications with relatively similar and moderate risk scores (around 2.5). This suggests a baseline consensus on their feasibility and the need for moderate regulation and oversight.

2. Bridge Gaps in Acceptance: By focusing on applications with relatively aligned views, developers can foster collaboration between fact-checkers and AI experts. This creates an opportunity to gain traction and build trust before tackling more divisive tools.

3. Address Emerging Needs: Early event verification and network-based analysis are crucial for fact-checking and disinformation monitoring, particularly in regions with high information warfare risks like Ukraine. These tools have strong practical applications and can improve response speed and accuracy.

Moreover, improving and building systems on the top of Image Reverse Search, Detection of Similar Fact-Checked Claims and Multimodal Verification, the tasks with high fact-checking acceptance, can also provide good avenues for trust improvement and open the floor for AI applications with more divisive views among fact-checkers and AI developers.

# IV. Metrics of success

The metrics of success for the proposed policies for media regulation:

- Reduction in the prevalence of dis- and misinformation cases based on independent audits.
- Average time to resolve cases of media negligence or intentional disinformation involving AI content and AI fact-checking tools.
- Compliance rates for mandated corrections or removals of false information.
- Public perception surveys on the fairness and transparency of regulatory bodies.
- The number of journalists and media professionals trained in AI literacy workshops.
- Participant satisfaction rates from training sessions.

The metrics of success for the proposed policies for European AI development:

- Performance metrics of state of art for AI tools for fact-checking, such as F1 score, MCC, and ECE, in detecting disinformation.
- The number of available benchmarks of different genres.
- The proportion of papers published on this topic at international conferences from European consortiums.
- Percentage of AI-generated content flagged as unverifiable or partially verified on the media.
- The proportion of AI tools audited and approved under the defined standards within media organisations in Ukraine and Germany.
- User feedback scores on tool usability and trustworthiness.

The metrics of success for the proposed policies for governments and funding bodies:

- Total funding allocated to projects promoting AI literacy and fact-checking.
- The number of funded projects completed successfully and their outcomes (e.g., publications, benchmarks), especially based on the percentage of tools produced by funded projects implemented into fact-checkers routine.

- The number of social media platforms and research projects joining the proposed working group consortium.
- The creation of the proposed Association of the European Social Media Research.
- The number of joint projects involving EU candidate countries and European Association, Candidate states and other international partners.
- The volume of shared resources (datasets, tools) produced from collaborative initiatives.
- The number of interdisciplinary AI and Computer Science programs established for policymakers.
- Enrollment rates and graduation outcomes of these programs.
- Surveys measuring trust levels in media, journalism, and government fact-checking efforts.
- Attendance and participation in public workshops and forums related to AI and media regulation

# V. Timeline

While AI development recommendations can be implemented as a comprehensive strategy without specific prioritization, media regulation and governmental funding recommendations can be conditionally categorized into short-, medium-, and long-term policies.

---

## Short-Term Steps (1-2 years)

**A. Media Regulation**

1. **Establish Legal and Structural Frameworks (Ukraine)**

   - Identify and adapt elements from the German State Media Treaty and EU Digital Services Act (DSA) to align with Ukrainian governance structures.
   - Draft and pass preliminary legislation to create independent media regulatory bodies at the regional and national levels.
   - Begin transparent public selection processes for regional and national regulatory committee members.

2. **Strengthen Disinformation Regulations (Germany)**

   - Organize public debates and surveys to initiate discussions on modernizing media laws in the context of digital information warfare.
   - Draft legislation focused on directly addressing disinformation instead of relying on secondary justifications (e.g., lack of a broadcasting license).

3. **Define AI Usage Standards in Media (Germany, Ukraine)**

   - Create guidelines limiting fully autonomous AI systems in journalism to media monitoring tasks.
   - Publish initial standards for "6-eye principle," bias reduction, and AI-generated content handling (e.g., "AI-verified" labels).
   - Fund experimentation hubs programs for media organisations.

**B. Funding Strategies**

1. **Promote AI Literacy and Digital Education (Germany, Ukraine)**

   - Launch AI literacy workshops for journalists, fact-checkers, and policymakers.
   - Start preparatory work for pilot programs for master's degrees in AI and digital journalism aimed at students with non-technical backgrounds.
   - Increase the number of 100% funded PhD positions in digital journalism.
   - E.g. via scholarships, start funding students with interdisciplinary backgrounds to do PhD in AI and Computer Science.

2. **Encourage Collaborative Projects with EU Candidate Countries (Germany, the EU)**

   - Initiate funding for smaller pilot collaborations involving Germany with Ukraine, Moldova, and Georgia to showcase the benefits of partnerships.
   - Expand funding calls for fact-checking and disinformation resilience projects under the BMBF and other EU-aligned bodies.

3. **Establish a Cross-Platform Data Accessibility Initiative consortium (Germany)**
   - Create a consortium involving media outlets, researchers, regulators, and platforms to address data-sharing challenges.
   - Issue appropriate funding.

---

# Medium-Term Steps (2-5 years)

**A. Media Regulation**

1. **Strengthen Media Oversight and Accountability (Ukraine)**

   - Fully operationalize independent regulatory bodies with regional offices and a central national committee.
   - Pass laws addressing intentional disinformation, including clear criteria for proving intent and third-party interests.

2. **Modernize disinformation legislation (Germany)**

   - Finalise legislative reforms balancing freedom of speech with disinformation regulation, informed by public consultations.
   - Establish stronger EU harmonization of disinformation laws through the Digital Services Act and related regulations.

3. **Expand AI Standards for Fact-Checking (Germany, Ukraine)**

   - Roll out updated AI standards emphasizing explainability, bias audits, and decision-making transparency in journalism.

**B. Funding Strategies**

1. **Expand Funding Opportunities (Germany)**

- Launch larger, multi-year funding programs for collaborative research between the EU and candidate countries.
- Increase the number of 100% funded PhD positions in digital journalism and interdisciplinary AI.

2. **Establish Cross-Platform Data Accessibility Initiatives (Germany, the EU)**

- The consortium establishes all scaleable frameworks and partnerships for the European Association of Social Media Research.
- Develop the first versions of updatable shared benchmarks and research APIs.
- First European Union-level funding is allocated to this project's scaling, connecting partners all over the EU.
- Jointly with the involved industry representatives fund and establish the proposed post-grad school for social media research between industry and academic.

---

## Long-Term Steps (5 years): Structural and Transformational Changes

**A. Media Regulation**

1. **Integrate Media Regulatory Frameworks into EU Structures (Ukraine, Germany, the EU)**

    - Align Ukraine's media regulations with EU standards as part of its European integration efforts.
    - Strengthen cross-border cooperation for tackling disinformation campaigns.

2. **Achieve Full Legislative Reform (Germany, the EU)**

    - Embed reformed laws into national governance and monitor their effectiveness over time.
    - Expand EU-wide cooperation to address evolving disinformation tactics.

**B. Funding Strategies**

1. **Create a European Association for Social Media Research (the EU)**

    - Form a permanent body to oversee cross-platform data-sharing, benchmark creation, and AI development collaboration.
    - Ensure the participation of candidate and associate countries to foster broader EU integration and resilience.

2. **Foster Global AI Collaboration (the EU)**

    - Build partnerships with international stakeholders to position Europe as a leader in social media disinformation research and fact-checking tools.
    - Scale up funding for interdisciplinary research and joint PhD programs between academia and industry.

# Conclusion

The challenges posed by disinformation and the evolving role of AI in journalism and fact-checking require a multifaceted response that balances innovation, regulation, and collaboration. This policy paper underscores the critical need for independent and transparent media regulatory bodies, robust funding mechanisms, and interdisciplinary training programs to bridge the gap between technology and governance.

In Ukraine, establishing a legal and structural framework tailored to its governance and societal needs is pivotal, drawing on successful models like Germany's Medienstaatsvertrag while adapting them to local contexts. Meanwhile, Germany could continue advancing its leadership within the European Union by supporting collaborative research with non-EU countries, particularly those facing serious disinformation challenges, such as Ukraine, Moldova, Georgia, North Macedonia, etc. These partnerships can foster innovation, share best practices, enhance common resilience against disinformation narratives and lay the groundwork for shared technological progress and mutual capacity building, also contributing to a more successful accession of the candidate states into the European Union.

Expanding governmental funding strategies is essential to support the advancement of AI tools and foster international collaboration. Investment in AI literacy and digital information programs at various educational and governmental levels is critical to empower key stakeholders with the skills necessary to effectively deploy and regulate AI tools. Funding of full PhD positions in digital journalism and AI specialisations would ensure the cultivation of a new generation of interdisciplinary experts who can bridge technical and policy divides.
To this end, fostering shared infrastructure such as centralised computing clusters and the proposed scalable consortium working towards establishing the European Association for Social Media Research. Its goal is to lay the foundation for easier research access to real-time social media data and community datasets that can alleviate the computational barriers and data scarcity that are the current main researcher bottlenecks for improving AI tools for disinformation monitoring. The recommended policy offers an equitable and environmentally conscious pathway for innovation, putting Europe at the centre and avant-garde of technological and regulatory progress and ensuring its better collective disinformation resilience.

AI tools designed for journalism and fact-checking must prioritise reliability, transparency, and scalability. By integrating explainable AI, fostering user trust through iterative feedback, and encouraging the development of computationally sustainable, specialised models, these technologies can better align with the nuanced demands of journalists and fact-checkers. Furthermore, addressing the data bottleneck through regulatory reforms and collaborative initiatives will unlock the full potential of AI solutions while ensuring ethical compliance.

Ultimately, this paper advocates for a future where AI acts as a powerful decision-support tool, complementing human expertise rather than replacing it. Achieving this vision requires coordinated efforts across policymakers, researchers, and industry leaders to create a transparent, adaptable, and resilient information ecosystem. Only through such collective action can we safeguard the integrity of journalism and uphold democratic values in the digital age.

# Appendix

## Interviewees

---

Germany: Media Representatives

**Caroline Lindekamp (CORRECTIV)**

Caroline Lindekamp is leading a BMBF-funded research project noFake: an interdisciplinary project, combatting disinformation with crowdsourcing AI-empowered platform for citizen journalists. Ms Lindekamp was a media analyst at NewsGuard, a university lecturer and researcher, and she is a journalist with a portfolio, including articles for Handelsblatt, Zeit Online, T3N and Journalist.

**Henrike Reintjes (RBB)**

Henrike Reintjes works as an Information Specialist in the Research and Information Service division of the Archive and Documentation department at Rundfunk Berlin-Brandenburg (RBB) in Berlin. She is part of the research service team in the Crossmedia News Centre and the department's innovation team.

**Rachel Baig (Deutsche Welle)**

Rachel Baig is an editor, journalist and media trainer at DW and DW Akademie, mainly in the news team. She is a founding member of DW's fact-checking team and presents the fact-checking format for YouTube in English.

**Christina Elmer (TU Dortmund)**

Prof. Dr Christina Elmer is an expert in data journalism, focusing on digital journalism processes, user-centred editorial product development, and the ethical dimensions of digital media. She held leadership roles at Der Spiegel, including Deputy Development Director and Head of Data Journalism, and has been a lecturer since 2007 for institutions like UdK Berlin, HAW Hamburg, and TU Dortmund. In her current consortium, she collaborates with a network of fact-checkers and institutions like CORRECTIV and the Austrian Institute of Technology aiming to integrate AI-based tools for identifying manipulated images and videos.

## Germany: AI developers

**Dorothea Kolossa (MTEC, TU Berlin)**

Prof. Dr. Ing Dorothea Kolossa is a distinguished researcher in the fields of signal processing, artificial intelligence, and cognitive systems, holding a W3 professorship at TU Berlin for Electronic Systems in Medical Technology. Prof. Kolossa is a coordinator of the noFake project in collaboration with the nonprofit investigative journalism newsroom CORRECTIV and scholars from Ruhr University Bochum (RUB) and Technical University (TU) Dortmund.

**Vera Schmidt (XplaiNLP/ QUL, TU Berlin)**

Dr Vera Schmitt leads the XplaiNLP group at the Quality and Usability Lab in TU Berlin, focusing on NLP, explainable AI, HCI, and the legal aspects of AI in disinformation detection and medical data processing in collaboration with various partners, including DFKI, RBB, DW, and Fraunhofer, funded by her successful acquisition of third-party projects. Dr. Schmidt also co-founded CorrelAid, contributing to international data science projects.

## Germany: Government representatives

**Hanna Katarina Müller**

Hanna Katharina Müller is the Head of the Division "Political Systems; Hybrid Threats; Disinformation" at the German Federal Ministry of the Interior and Community (BMI). Since joining the ministry in January 2019, she has led efforts to counter disinformation and hybrid threats, including the establishment of a task force against disinformation and plans for a citizens' council in 2024. In June 2024, she also became the head of the development team for the newly established "Central Office for the Detection of Foreign Information Manipulation" (ZEAM), aiming to quickly detect and analyse foreign information manipulation campaigns.

**Benedict Föll**

Benedict Föll is a lawyer in the legal department and regulations department of the federal states media regulator for Berlin Brandenburg, which overviews online media and public broadcasting in these federal states, also funding local journalistic projects for information and news literacy.

## Ukraine: Media representatives

**Valeriia Stepaniuk (VoxUkraine)**

Valeriia Stepaniuk is a Deputy Director of VoxCheck and was previously an analyst at VoxCheck. She graduated from the Institute of International Relations at Taras Shevchenko National University of Kyiv, specializing in international relations.

**Yehor Brailian (Detector Media)**

Dr Yehor Brailian is a Ukrainian historian and journalist, an active contributor to media on foreign affairs. His work has appeared in outlets like Ukrinform, European Pravda, HVG, and the Wilson

Center blog and comments for BBC, Radio Free Europe, Espreso TV, Suspilne, BelSat on international relations. He has been an analyst at Detector Media since 2023, focusing on Russian disinformation and its global impact.

## Ukraine: Ukrainian AI developers

**Viktoria Skorik (Mantis Analytics)**

Viktoria Skorik is a Founding ML engineer at Mantis Analytics, working on disinformation detection and text and image processing algorithms. She previously worked at SoftServe, where she specialised in deep learning, data mining, and time series forecasting, while she also holds a Master's degree in Artificial Intelligence Systems and a Bachelor's degree in Computer Science from Kharkiv National University of Radioelectronics.

**Kostiantyn Zahorulko (Mantis Analytics)**

Kostiantyn Zahorulko is a Data Scientist with over five years of experience in data analytics and modelling, specializing in the finance, banking, and digital marketing sectors. He holds a Master's degree in Economic Analysis from the KSE (Kyiv School of Economics). Kostiantyn worked at such organizations, as KSE, PrivatBank, Together Networks, and SoftServe and is currently a Machine Learning Engineer at Mantis Analytics.

## Ukraine: Government Representatives

**Dmytro Zolotuhin**

Former Deputy Minister of Information Policy of Ukraine (2017–2019) and expert in information warfare and competitive intelligence. Since 2003, he has worked in Ukraine's national security bodies. As Deputy Minister, he oversaw information security, the development of Ukraine's strategic communications system, and the implementation of the Information Security Doctrine, which he co-authored.

**Anastasia Ostrovska**

Head of Communications of the National Coordination Centre for Cybersecurity at the National Security and Defence Council of Ukraine. Communications Advisor to Vice Prime Minister of Ukraine (2020-2021), co-founder of Kyiv International Cyber Resilience Forum.